# Collective electronic excitation in a trapped ensemble of photogenerated dipolar excitons and free holes revealed by inelastic light scattering

Sebastian Dietl[1,2], Sheng Wang[3], Dieter Schuh[5], Werner Wegscheider[6], Jörg P. Kotthaus[4], Aron Pinczuk[3,7], Alexander W. Holleitner[1,2,*], Ursula Wurstbauer[1,2,*]

[1]*Walter Schottky Institut und Physik Department, Am Coulombwall 4a, Technische Universität München, D-85748 Garching, Germany*

[2]*Nanosystems Initiative Munich (NIM), Schellingstr. 4, 80799 München, Germany*

[3]*Department of Applied Physics and Applied Mathematics, Columbia University, New York, New York 10027, USA*

[4]*Center for Nanoscience and Fakultät für Physik, Ludwig-Maximilians-Universität, Geschwister-Scholl-Platz 1, 80539 München, Germany*

[5]*Institute of Experimental and Applied Physics, University of Regensburg, D-93040 Regensburg, Germany*

[6]*Solid State Physics Laboratory, ETH Zurich, 8093 Zurich, Switzerland*

[7]*Department of Physics, Columbia University, New York, New York 10027, USA*

Abstract

Photogenerated excitonic ensembles confined in coupled GaAs quantum wells are probed by a complementary approach of emission spectroscopy and resonant inelastic light scattering. Lateral electrostatic trap geometries are used to create dense systems of spatially indirect excitons and excess holes with similar densities in the order of $10^{11}$ cm$^{-2}$. Inelastic light scattering spectra reveal a very sharp low-lying collective mode that is identified at an energy of 0.44 meV and a FWHM of only ~50 µeV. This mode is interpreted as a plasmon excitation of the excess hole system coupled to the photogenerated indirect excitons. The emission energy of the indirect excitons shifts under the application of a perpendicular applied electric field with the quantum-confined Stark effect unperturbed from the presence of free charge carriers. Our results illustrate the potential of studying low-lying collective excitations in photogenerated exciton systems to explore the many-body phase diagram, related phase transitions, and interaction physics.

*corresponding authors: holleitner@wsi.tum.de and wurstbauer@wsi.tum.de

During the last decades, there have been considerable efforts to realize and study Bose-Einstein condensation (BEC) of interacting bosonic particles by creating excitons in solid state systems [1]. BEC has been observed for the first time in atomic systems [2,3] and later also in more exotic solid-state systems like exciton-polaritons [4–6]. Already more than 50 years ago, BEC has been predicted for excitons in semiconductors [7]. Most of such experiments are based on two tunnel-coupled semiconductor quantum wells (CQWs) hosting the excitons. These are electron-hole pairs coupled by the attractive Coulomb interaction. The bosonic quasiparticles exhibit a rich quantum phase diagram including a free exciton gas at elevated temperatures and low densities, which is expected to condense into a BEC with a macroscopic coherence at low temperatures. At high densities, an electron-hole plasma is predicted at higher temperatures that undergoes a Bardeen-Cooper-Schrieffer (BCS) transition to an excitonic insulator at low temperatures [1]. Also, signatures for a (classical) exciton liquid formed by repulsive exciton interactions have been reported [8]. The phase transition between the different phases might be gradual, and the coexistence of different (classical and quantum) phases is possible. Each phase inherently holds a characteristic spectrum of low energy collective excitations of strongly or weakly coupled electron-hole pairs including phenomena such as roton instabilities in their wave vector dispersion [9,10]. In this context, it has been shown for electron bilayers at a total filling factor of one, where exciton condensation occurs [11–14], that deep and soft magnetorotons as well as collective spin excitations exist [15,16]. These excitations are linked to quantum phase transitions [15,16].

In general, there are two types of CQW systems hosting excitons. In one case, both CQWs are doped with electrons, and a strong perpendicular magnetic field is applied such that in each quantum well, the lowest Landau level is exactly half filled with electrons and consequently also half filled with holes. If the Coulomb interaction is strong enough in such CQWs, the electrons and holes in the two CQWs can form metastable excitons. Such systems have been realized for instance in GaAs CQWs [11–14] but also very recently in double graphene bilayers [17] and in a hybrid system of GaAs and graphene [18]. Experimental signatures for the collective behavior in such systems include exciton superfluid phases, exciton condensation and Coulomb drag effects [13], but also collective excitations such as roton minima [15] similar to superfluid helium [19,20] as well as spin excitations [15,16]. In a second case of a CQW system, the electron hole pairs are optically excited and then spatially separated by the application of a perpendicular electric field such that the electrons tunnel to one quantum well and the holes to the other one. Such indirect excitons (IXs) exhibit long photoluminescence lifetimes up to several microseconds [21] and a permanent dipole moment given by the spatial separation of the photogenerated electrons and holes. Therefore, such IXs are also called dipolar excitons and are the subject of this report. By confining them in an electrostatic trap, one can create dense ensembles with the IX temperature close to the lattice temperature [22,23]. The transition to a BEC is supposed to occur below a certain critical temperature $T_c$, which is in a first approximation proportional to $n_{IX}/m^*$, with $n_{IX}$ the exciton density and $m^*$ the effective mass of the IXs [24]. The critical temperature for condensation in dipolar exciton systems can be estimated to be in the order of $T_c \approx 1K$ [25]. Dipolar exciton systems realized in GaAs and InGaAs based CQWs exhibit a variety of exciting experimental signatures [8,26–30]. Nevertheless, an unambiguous proof for BEC and long range coherence is challenging [27,31–33]. Hereby, the complex excitonic phase diagram including the nature of the transition between the different phases still needs to be explored in more detail. Most studies of IX ensembles rely on photoluminescence (PL) experiments. However, PL is very sensitive to any localization sites such as crystal imperfections, background doping and potential fluctuations masking the PL signal from the IX system. Furthermore, only excitons with total spin of +/-1 couple to light. Excitons with a total spin of +/- 2 form dark states. Since bright and dark states are expected to weakly couple [34] with an unknown coupling strength, PL can only probe part of the IX system.

Here, we introduce a combination of PL measurements and resonant inelastic light scattering (RILS) on elementary collective excitations of the electron-hole system to study dense IX ensemble. RILS is proven to be a very powerful spectroscopy method to investigate collective modes of photogenerated electron-hole plasma in GaAs [35], to study charge and spin excitations as well as to explore quantum phase transitions and the nature of quantum phases in a variety of quantum systems such as electron bilayers [15,16], quantum Hall [36] and fractional quantum Hall effect systems [37,38]. Furthermore, collective excitations in quantum wires [39–41], quantum dots [42,43] in dilute two-dimensional electron systems [44] and the spin-splitting in the hole dispersion [45] have been studied by RILS. Generally, collective excitations of the photogenerated exciton ensembles are expected to serve as a unique fingerprint to unambiguously identify the individual states in the phase diagram of dipolar exciton ensembles. In this way, the investigation of collective excitations should allow the monitoring of transitions between such phases. Studying collective excitations is commonly less sensitive to disorder compared to PL since the collective modes' energy is unperturbed and only slightly broadened by moderate residual-disorder. With our complementary approach, we find that at elevated temperatures of about 7 K and high excitation intensities, the photogenerated IX ensemble coexist with excess holes of similar densities. A low-lying collective mode is observed in the RILS measurements. This mode is identified as a plasmon excitation of the photogenerated two-dimensional (2D) excess holes. Compelling evidence for the many-body interactions between the IXs and 2D hole system is found in the measured plasmon energy that is well described by an effective screening of the plasmon mode within the CQWs [46,47].

The photogenerated exciton ensembles are hosted by a GaAs-based heterostructure grown by molecular beam epitaxy on semi-insulating (001) GaAs substrates. The heterostructure consists of two 8 nm wide GaAs quantum wells separated by a 4 nm $Al_{0.3}Ga_{0.7}As$ tunnel barrier. The CQWs are grown in a 370 nm thick field-effect layer with a heavily *n*-doped GaAs back-gate electrode and 6 nm thick semi-transparent top-gate electrodes made from titanium. The distance from the center of the upper quantum well to the top (epitaxial back) gate is 54 nm (316 nm). As sketched in Fig. 1(a), a circular top electrode with a diameter of 23 µm (trap) is surrounded by a second top electrode (guard). Both gates are biased with respect to the back gate enabling a precisely defined lateral potential landscape for the IXs [48,49]. The energy of the IXs with respect to the direct excitons (DX) energy is shifted by $-\vec{p} \cdot \vec{F}$, where $\vec{p}$ denotes the effective exciton dipole moment and $\vec{F}$ the electric field amplitude. The electric field can be estimated to be $\vec{F}$ = - ($V_g$ + $V_s$)/d, with $V_g$ the voltage between bottom and top gate electrodes and $V_s$ = -0.7 *V* the effective height of the metal Schottky barrier. The distance between the top and bottom gates is *d* = 370 nm. Photogenerated IX ensembles with densities in the order of $n_{IX}$ ~ $10^{11}$ cm$^{-2}$ can be defined under the trap gate for |$F_{trap}$| > |$F_{guard}$| [50]. In this situation, the potential landscape is such that electrons can effectively escape from the trap region while the less mobile holes are trapped resulting in an excess 2D hole system in the trap region with 2D hole densities $n_h$ in the order of those of the IX system [51]. The densities of the IXs and holes depend on the applied electric field and the illumination intensity. The IX density $n_{IX}$ can be determined from the blue-shift of the emitted PL light. The two-dimensional hole density $n_h$ is more difficult to be estimated from experiment, but it is known for the device under investigation from Ref. [50], where the authors have extracted $n_h$ with the help of Landau level spectroscopy.

The sample is placed in a backscattering geometry with a finite tilt angle of $\theta$ = 32° ± 2° in a $^4$He bath cryostat slightly above the liquid He level [Fig. 1(c)] and hence, only cooled by He vapor to a temperature of ~7 K. The light of a tunable titanium:sapphire laser with a FWHM ≤ 20 µeV is focused to a 20 µm spot within the trap gate on the sample surface by a combination of a beam expander and a lens. The emitted as well as scattered light is collected and focused on the entrance slit of a double grating spectrometer by a lens system. The tilt between the sample normal and the incident (scattered) light, that is shown in inset of Fig. 2(a), allows the in-plane transfer of a finite momentum $q$ = |$k_{laser}$ - $k_S$| = 4π/λ·sinθ = 2$ω_{laser}$/c·sinθ ≈ 8.6 · $10^4$ cm$^{-1}$ to the optically excited system coexisting of IXs and free holes. The $k_{laser(S)}$ are the wave vectors of the incident laser and scattered light, *λ* ($ω_{laser}$) is the wavelength (frequency) of the exciting laser light, and *c* the speed of light in vacuum. In RILS, the energy shift is $E_{RILS}$ = *ΔE* = ℏ·|$ω_{laser}$ – $ω_s$|, where $ω_S$ is the frequency of the scattered light. To minimize the reflected and Rayleigh-scattered light

at the energy of the exciting laser at $\Delta E = 0$, the RILS measurements are in the so-called depolarized geometry with the linear polarizations of the incident and scattered light perpendicular to each other.

Figure 1(b) shows a photoluminescence spectrum taken at a temperature of 7 K and excited with $\hbar\omega_{laser}$ = 1.6090 eV, a light intensity of 8.5 W/cm$^2$ for an electric field configuration of $F_{trap}$ = 27.0 kV/cm under the trap and $F_{guard}$ = 13.5 kV/cm under the guard region. The experimental parameters correspond to an IX density of $n_{IX} \approx 7.0 \times 10^{10}$ cm$^{-2}$. The IX density is deduced from the measured photoluminescence blue-shift, taking into account a density and temperature dependent correction term $f(n,T) \approx 0.46$ [23]. The free hole density is reported to $n_h \approx 2 \times 10^{11}$ cm$^2$ for the same device at the given experimental conditions [50]. We observe photoluminescence from the IX system under the trap electrode at $E_{IX}^T$ = 1.5489 eV, from the IX system under the guard electrode at $E_{IX}^G$ = 1.5596 eV, and from the direct excitons at $E_{DX}$ = 1.5751 eV [Fig. 1(b)]. We simultaneously detect the emission from both the trap and the guard gate region in the spectrum, since the spot diameter of approx. 20 µm is as large as the trap gate. Due to mechanical vibrations of the cold finger in our continuous wave measurements, both trap and guard regions are excited on the samples. The different electric fields cause different shifts of the IX systems and enable to clearly separate the emission from the two gate regions.

In a next step, RILS spectra are taken with similar electric fields at the guard and trap gates as for the emission spectra displayed in Fig. 1(b). We observe a well-defined low-energy mode in the RILS spectra at $E_{RILS}$ = 0.44 meV that is resonantly enhanced. The results in Fig. 2(a) show such a RILS spectrum for $\hbar\omega_{laser}$ = 1.59691 eV and a light intensity of 110 W/cm$^2$. The background signal increases towards lower energies and it is caused by elastically scattered Rayleigh light centered at $\Delta E$ = 0 meV that is well reproduced by a single Lorentzian profile. Fig. 2(b) shows the background corrected RILS spectra taken at different laser energies in a narrow energy range of only $\Delta\hbar\omega_{laser}$ = 0.58 meV. The low energy mode is resonantly enhanced within this narrow range of exciting photon energies. Moreover, the peak is very narrow identifying the mode as a collective electronic excitation, e.g. a plasmon mode. The RILS mode is maximally enhanced for $\hbar\omega_{laser}$ = 1.59691 eV, which is significantly above the emission energy of the DXs at $E_{DX}$ = 1.5751 eV. We note that RILS measurements close to $E_{DX}$ are very difficult because of the strong PL background at this energy. Instead, Fig. 2 demonstrates that there is an enhanced RILS mode at an interband transition, which is not the transition between conduction band minimum (*e*) and the maximum of the heavy hole valence subband (*hh*) in the QWs. Below these two single particle bands, an exciton with 1s-symmetry is formed (DX emission line) with an exciton binding energy below 10 meV [52].

To determine the higher lying interband transition causing the resonant enhancement of the RILS mode, we exploit the leakage photocurrent between the trap- and the back-gate electrodes [53–55] in combination with numerical band structure calculation for the CQWs with a perpendicular applied electric field of $F$ = 22.7 kV/cm (Fig. 3). The photocurrent is directly proportional to the light absorption in the CQWs structure and therefore, it enables the determination of optical transitions also from higher lying interband transitions. Figure 3(a) displays the corresponding normalized photocurrent spectrum versus photon energy. The measurements are carried out for an applied voltage at the trap-gate electrode of $V_{trap}$ = -0.4 V resulting in an electric field of $F$ = 29.7 kV/cm, whereas the voltage between guard-gate and back-gate electrode is kept at $V_{guard}$ = -0.1 V. We detect two pronounced maxima in the photocurrent and hence, in the absorption of the CQWs structure. We estimate the peak positions to be at $I^{PC}_{e-hh}$ ≈ 1.5784 eV and $I^{PC}_{e-lh}$ ≈ 1.5973 eV. We attribute them to the optical transitions from the heavy hole (*hh*) and light hole (*lh*) subbands to the lowest electronic subband (*e*) in the QWs. In particular, the lowest energy (*e-hh*) transition in the photocurrent measurement is slightly above but in reasonable agreement with the DX emission peak at $E_{DX}$ = 1.5751 eV [compare Fig. 1(b)]. The higher energetic peak in the photocurrent spectrum marked with an arrow in Fig. 3(a) agrees well with the incoming photon energy for which the mode in RILS is resonantly enhanced. This photocurrent peak is assigned to a transition between the lowest conduction band (*e*) with the first excited valence-band in the QW with light-hole character (*lh*). We corroborate this assumption with numerical simulations of the single particle band structure in the CQWs with an applied electric field performed with nextnano$^3$ [56]. The calculations are performed without considering Coulomb interactions and therefore without the exciton binding energy. We calculate the eigenstate wave functions and the corresponding energies in the CQWs in an 8 x 8 $k \cdot p$ approximation. The results are summarized in Fig. 3(b), where the relevant transitions between *e-hh* and *e-lh* are marked by arrows. The superscripts top (t) or bottom (b) indicate whether the states are localized in the bottom or top quantum well, respectively. The absolute transition energy extracted from the calculations differs only by a few meV from the DX emission energy. This could be due to the fact that the calculations are done without exciton binding energy, due to minor variation in the heterostructure or minor uncertainties in the applied electric field. The energy difference between heavy hole (*hh*) and light hole (*lh*) transitions with the lowest conduction band, however, agrees very well with the experimental data observed from photocurrent spectroscopy and corroborate the above given interpretation that the lower energy peak in the photocurrent measurements stems from an absorption at the *e-hh* transition, whereas the higher energy peak stems from the *e-lh* transition. Consequently, the collective electronic mode exhibits a clear enhancement for excitation with an energy belonging to the *e-lh* transition.

Before turning to the discussion of how to interpret the observed collective electronic excitations in RILS, we would like to confirm the excitonic functionality of our trap under the utilized laser power and bath temperature. The response of the trapped IX ensemble in photoluminescence measurements is analyzed vs. the trap gate voltage/field at constant guard $F_{guard}$ = 13.5 kV/cm [Fig. 4(a)]. As expected, the emission energy of the IXs shifts according to the linear quantum-confined Stark effect with (1.14 ± 0.03) meV/(kV/cm). The value is reproduced by theoretical calculations with nextnano[3], and it confirms that, despite an eventual tunneling current along the growth direction, there seems to be no substantial screening of the applied electric field across the device. The slight deviation from the linear behavior at $V_{trap}$ = 0.3 V ($F_{trap}$ = 10.8 kV/cm) is due to the density increase at the trap – antitrap transition induced by the guard electrode [51]. In this context, the 'trap configuration' refers to $V_{trap}$ < $V_{guard}$, and the 'antitrap configuration' refers to $V_{trap}$ > $V_{guard}$, respectively.

It has been demonstrated by K. Kowalik-Seidl and co-workers [50], that the voltage regime affects the escape dynamics of unbound charge carriers from the trap. Particularly, in the trap configuration, the potential landscape below the electrodes allows unbound electrons to escape from the trap, whereas unbound holes remain confined below the trap electrode. The coexistence of such trapped, excess holes with density $n_h$ is further driven by high excitation powers and elevated temperatures, as has been quantified by Landau Level spectroscopy [50]. Again, the density of the IXs can be estimated by measuring the interaction-induced blue-shift of the IX emission taking into account the correction factor *f(n,T)* [23]. Although the exact correlation between exciton density and photoluminescence blue-shift remains controversial, it has been mostly accepted that the interaction energy can be expressed as [57]:

$$E_{int} = \frac{4\pi e^2 d}{\varepsilon} n_{IX} \cdot f(n,T), \qquad (1)$$

with *d* = 12 nm the separation between the QW center, *ε* = 12.9 the permeability of GaAs, $n_{IX}$ the indirect exciton density and *f(n,T)* a correction factor taking into account the effective exciton interaction potentials [23]. In order to determine the maximum exciton density in our experiment, we measure the IX photoluminescence energy as a function of incident power density [Fig. 4(b)]. We note that this data set is taken in a different, fiber-baser optical setup with superior optical resolution of about 1 μm to ensure the excitation only within the trap region. We measure a total blue-shift of 11.33 meV, when the excitation power density is varied between 0.13 W/cm² and 124 W/cm². This corresponds to an increase in density from $n_{IX} \approx 1.1 \times 10^{11}/cm^2$ to $n_{IX} \approx 6.7 \times 10^{11}/cm^2$, depending on the choice of $f(n,T)$ (0.1 ≤ $f(n,T)$ ≤ 0.6). The corresponding power-dependent photoluminescence spectra are shown in the inset of Fig. 4(b). The IX emission peak broadens with increasing power, but we do not observe any indication for recombination of free electrons and holes, e.g. an indication for crossing the so-called exciton Mott-transition [58]. We conclude that the system is still in the state of a free exciton gas. A closer look to the line shape of the emission peaks reveal for the highest excitation density that the emission peak is very well reproduced by a single Gaussian, whereas for the lowest excitation power, the emission peak is well reproduced by two Gaussians separated by ~1.5 meV as demonstrated in the inset of Fig. 4(b). Due to the single peak representation at the highest excitation power, we exclude the formation of charged IX (trions) in this regime, whereas the double peak structure in the low power regime might be a hint for the formation of charged excitons complexes [59].

To reveal the origin of the collective excitation mode, we take into account both subsystems in the trap, i.e. the IXs and the excess holes. Even though the photoexcitation for the RILS experiments also covers the guard electrode, the IX density there is much lower due to a lower electric field, a much lesser number of excited excitons and the missing lateral confinement. The latter allows the diffusion of the IXs away from the excitation spot. Similarly, the escaped electrons from the trap region are not laterally confined below the guard gate. They are expected to diffuse away from the excitation spot resulting in a vanishing electron density. As a result, the RILS mode is only reasonable due to a collective excitation of the ensemble under the trap: either collective hole excitations, such as plasmons, or a collective excitation of the IX system. For the latter case, an exciton liquid is expected to hold collective excitations with one or several critical points in their wave-vector dispersion such as roton minima and maxima at finite momenta [9,10] similar to the roton minimum in the excitation spectrum of liquid helium [19]. Such critical points with a high density of states are observable in RILS even if the transferred momentum $q$ of the photons is much smaller due to break-down of wave-vector conservation by residual disorder [38,60]. In our understanding, however, such an interpretation is unlikely since at a bath temperature of $T = 7$ K, we expect an exciton gas and not an exciton liquid [1]. Having this in mind, the only subsystem responsible for the collective mode is the confined 2D gas of excess holes.

Due to the narrow line width of only ~50 µeV of the measured RILS mode, we exclude single-particle-like excitations between the spin-split hole dispersion at finite momenta [45] as a feasible explanation of the observed mode. Instead, we interpret the RILS resonance in Fig. 1 as a 2D hole plasmon [61,62] because of the following reasons. The well-known wave-vector dispersion for plasma oscillations in 2D charge carrier systems is [46]:

$$\omega_p = \sqrt{(2\pi n_h e^2/\varepsilon^* m_{hh}^*)q}, \qquad (2)$$

with the effective permeability $\varepsilon^*$, hole density $n_h$ and effective hole mass $m^*_{hh}$. The values constitute to $n_h$ = 2 x 10$^{11}$ cm$^{-2}$ and $m^*_{hh}$ = 0.36 m$_e$, respectively. They are determined from Landau level spectroscopy as in Ref. [50]. With the value for $\varepsilon^*$ = $\varepsilon_{GaAs}$ = 12.9, eq. (2) gives $\hbar\omega_p \approx$ 1.6 meV, which significantly exceeds the energy of the measured RILS mode at $E_{RILS}$ = 0.44 meV by more than 1 meV. This energy difference is beyond the conjectured error in theoretical and measured values. We emphasize that the theoretical value for a plasma excitation of a 2D electron (and not hole) system would be even larger due to the lower effective mass of the electrons compared to the holes in GaAs-based QWs. Importantly, eq. (2) neglects the fact that the hole plasmon is screened by the free charge carriers available either in the nearby top and bottom gate electrodes as well as the photogenerated charge carriers within the CQWs, as theoretically expected for our heterostructure [47,63,64]. The device geometry with all important distances is sketched in Fig. 5. Taking into account a perfect screening by a metallic gate electrode and neglecting retardation effects, the effective permeability is given by [47,65]

$$\varepsilon^* = \varepsilon_1 \coth qd_1 + \varepsilon_2 \coth qd_2, \qquad (3)$$

where $\varepsilon_1$ and $d_1$ are the permeability and thickness of semiconducting layers between the upper QW and the top gate electrode with $\varepsilon_1$ = $\varepsilon_{GaAs}$ = 12.9 and $d_1$ = 54 nm. For the bottom gate, we get values of $\varepsilon_2$ = $\varepsilon_{GaAs}$ = 12.9 and $d_2$ = 316 nm, respectively. In this evaluation, we neglect the small difference in the permeability of GaAs and AlGaAs in the regions between QW and top or bottom gate electrode.

Equation (3) indicates an effective permeability $\varepsilon^*$ = $\varepsilon_{GaAs}$·(2.31 + 1.01) = $\varepsilon_{GaAs}$·3.32. In combination with eq.(2), the corresponding value for the plasmon energy is $\hbar\omega_p \approx$ 0.88 meV. This value is still more than 0.4 meV larger than the measured value of $E_{RILS}$ = 0.44 meV. When we assume that the hole plasmon is screened within the CQWs themselves, we calculate the 2D plasmon energy to be $\hbar\omega_p$(12 nm) $\approx$ 0.49 meV for an effective center distance between the electrons and holes in the two QWs to be $d_2$ = 12 nm, respectively. This theoretical estimate is in good agreement with the measured value of $E_{RILS}$ = 0.44 meV. In such an interpretation, the electrons residing in the lower QW forming the dipolar excitons [ellipses in Fig. 5] screen the 2D hole plasmon in the upper well. When we just take the width of the AlGaAs tunnel barrier within the CQWs, the distance $d_2$ in eq.(3) is further reduced to 4 nm. However, the calculated plasmon energy within this approach is $\hbar\omega_p \approx$ 0.29 meV, which is well below the experimental value. We should keep in mind that in all calculations, we neglect the effect of the exciton binding energy between the electrons in the lower QW with a fraction of the holes in the upper quantum well. Moreover, the repulsive Coulomb interaction in the 2D hole system in the upper QW is neglected. Both phenomena are expected to further modify the 2D hole plasma frequency in addition to the screening effect by the electrons in the lower QW. An interesting consequence of this interpretation is that also the dipolar exciton ensemble behaves collectively in an oscillating motion. In other words, the hole and the exciton systems are two coupled oscillator systems. In this respect, the presented RILS measurements reveal the combined dynamic susceptibility induced from the collective excitations of the 2D hole and dipolar exciton system. Along this line, the significant reduction of the 2D hole plasmon energy is tentatively seen as a manifestation of correlations between the IXs and the free hole subsystems effectively modifying the plasmon properties. We would like to note that such a deviation of the plasmon dispersion has also been observed for ultra-low density 2D electron systems. There, it was explained by correlation effects that effectively reduce the Coulomb interactions [44].

We note that if the polarization selection rules are strictly fulfilled, plasmon excitations should be only observable in the polarized scattering geometry with incoming and scattered light co-polarized [66,67]. In our experiments, however, the selection rules are less strict due to the combined effects of excitations under extreme resonance conditions [68], valence band mixing of heavy and light hole bands resulting in a band non-parabolicity [69] and an applied electric field perpendicular to the QW plane. Hereby, collective charge density modulation of the dynamic susceptibility, e.g. by a plasma excitation, can be observed in RILS measurements also in the depolarized geometry. Our results are promising for more detailed RILS experiments. In particular, we demonstrate that the optical transition from the light-hole to the lowest conduction band can be exploited to perform RILS experiments on IXs in CQWs. An exciting laser energy closer or equal to the direct exciton line gives rise to a dominating PL background. Hereby, the presented RILS measurements show a first route to access photogenerated IXs. Future experiments can be combined with coherence studies on trapped IXs with the aim to explore the complex density-temperature phase diagram in depth with special emphasize on the individual phase transitions. RILS experiments should be particularly suited to do so because the individual phases are unambiguously characterized by the collective electronic excitations. The underlying wave-vector dispersions are unique and they can be rather complex with roton minima und maxima.

In summary, the optical response of a photogenerated, laterally trapped ensemble of IXs in co-existence with an excess 2D hole system with similar densities is studied by a combination of photoluminescence emission, RILS, and photocurrent experiments. The emission spectra reveal that indirect excitons localized in the CQWs respond to the perpendicular applied electric field as expected for the quantum-confined Stark effect. In the RILS measurement, we find a sharp low energy mode at $E_{RILS}$ = 0.44 meV with a FWHM of only 50 µeV. The RILS mode is resonantly enhanced for an energy interband transition between the lowest conduction band and the first excited hole subband with *lh* character. The high absorption at this optical transition is demonstrated by the photocurrent experiments in combination with band structure calculation utilizing nextnano$^3$ [56]. The RILS mode is interpreted as a plasmon excitation of the 2D hole sub-system. The deduced plasmon energy is assigned to the many-body interaction of the two photogenerated subsystems, namely the IX gas and the excess 2D hole system. The presented experiments notably proof that the approach of combining RILS measurements on electronic excitations is a highly capable method to explore the phase diagram of optically excited IXs. We expect that by further cooling down the samples, the correlation between the subsystems increases as soon as the excitonic subsystems become a liquid. In particular, we expect to observe collective excitations of the IX liquid. Particularly roton excitations are expected to play a key role in experiments on trapped dipolar Bose-Einstein condensates [9]. Our complementary experimental approach can also be applied for CQWs made out of graphene bilayers [17] and vertical van der Waals heterostructures assembled by $WSe_2$ and $MoS_2$ e.g. with hBN as tunnel barrier [70].

Acknowledgement

We thank B. N. Rimpfl, X. Vögele und K. Kowalik-Seidl for sample preparation and initial sample characterization. The work is supported by the Deutsche Forschungsgemeinschaft (DFG) via excellence cluster 'Nanosystems Initiative Munich' as well as DFG projects WU 637/4-1 and HO3324/9-1. The work at Columbia is supported by the US National Science Foundation Division of Materials Research (Award DMR-1306976) and by the US Department of Energy Office of Science, Division of Materials Science and Engineering (Award DE-SC0010695).

[62] A. S. Plaut, A. Pinczuk, B. S. Dennis, C. F. Hirjibehedin, L. N. Pfeiffer, K. W. West, *Appl. Phys. Lett.* **2004**, *85*, 5625.
[63] A. V. Chaplik, *Soviet Physics JETP* **1972**, *35*, 395.
[64] F. Stern, *Phys. Rev. Lett.* **1967**, *18*, 546.
[65] D. A. Dahl, L. J. Sham, *Phys. Rev. B* **1977**, *16*, 651.
[66] S. Das Sarma, D.-W. Wang, *Phys. Rev. Lett.* **1999**, *83*, 816.
[67] B. Jusserand, M. N. Vijayaraghavan, F. Laruelle, A. Cavanna, Etienne B., *Phys. Rev. Lett.* **2000**, *85*, 5032.
[68] A. Pinczuk, L. Brillson, E. Burstein, E. Anastassakis, *Phys. Rev. Lett.* **1971**, *27*, 317.
[69] R. Winkler, *Spin-Orbit Coupling Effects in Two-Dimensional Electron and Hole Systems*, Springer Berlin Heidelberg **2003**.
[70] M. M. Fogler, L. V. Butov, K. S. Novoselov, *Nature communications* **2014**, *5*, 4555.
14

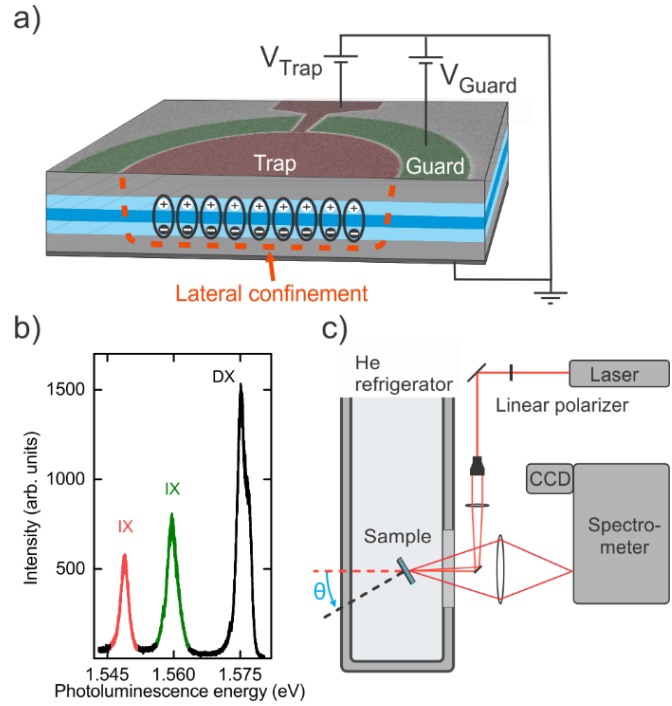

*Figure 1: (color online) (a) Schematic view of the coupled quantum well field-effect trap device in cross-section, covered by a projection of a scanning electron microscope (SEM) image of the center trap (guard) electrode. The trap diameter is 23 µm. (b) Photoluminescence (PL) spectrum of direct (DX) and indirect (IX) excitons measured for T = 7 K, electric field strength $F_{trap}$= 27.0 kV/cm, $F_{guard}$= 13.5 kV/cm and incident laser power density of 8.5 W/cm² at a photon energy of 1.609 eV. The emission at ~1.5489 eV (~1.5596 eV) stems from the region of the CQW below the trap (guard) gate. The experimental setup is sketched in (c). The sample is placed slightly above the liquid $^4$He level inside a cryostat in a back scattering geometry.*

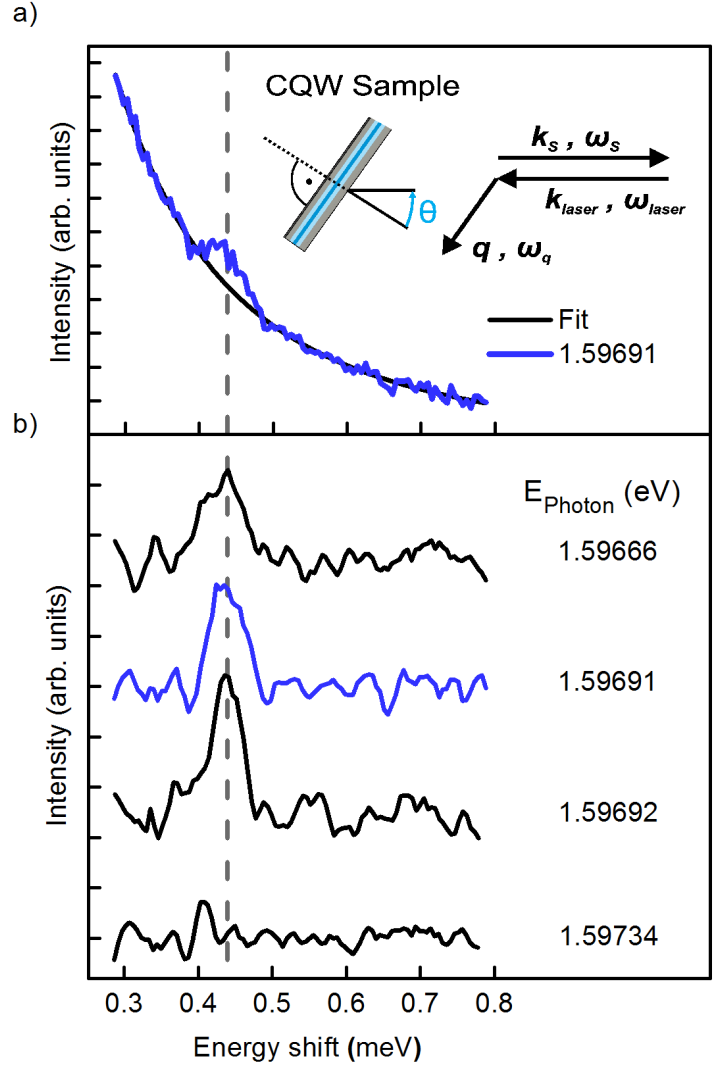

Figure 2: (color online) (a) RILS spectrum and Lorentz-fit of the laser background signal at $F_{trap}$ = 24.3 kV/cm and at $F_{guard}$ = 18.9 kV/cm. Inset: Sketch of the backscattering geometry. $k_{laser}$ ($k_S$) and $\omega_{laser}$ ($\omega_S$) denote the wave vector and angular frequency of the incident (backscattered) beam. The transferred wave vector and frequency are $q$ and $\omega_q$, respectively. (b) RILS spectra for different incident photon energies. A background is subtracted and the data are smoothed. The second from the top spectrum is shown as raw data in (a). The dashed line is a guide to the eye.

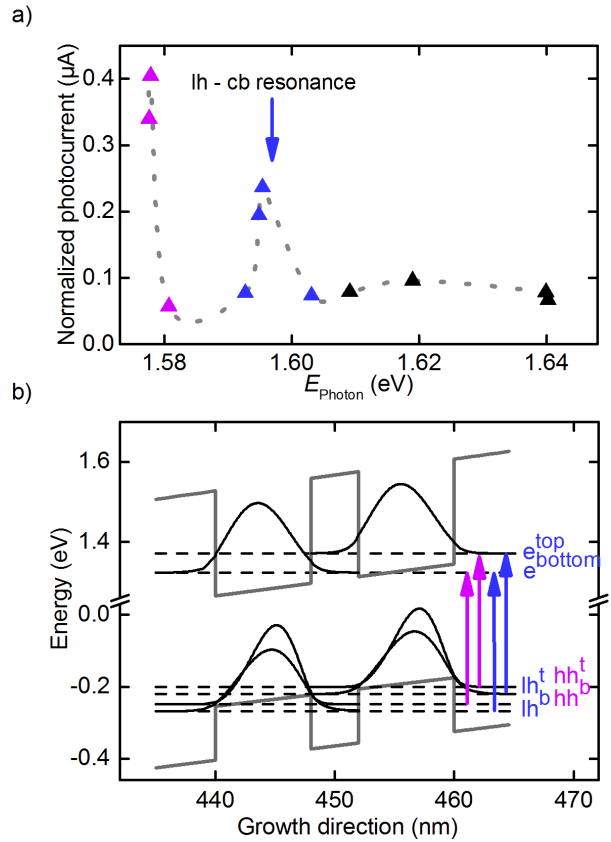

Figure 3: (color online) (a) Photocurrent perpendicular through the heterostructure under illumination, normalized to the illumination power (T = 7 K, P = 96 – 176 W/cm², $F_{trap}$ = 29.7 kV/cm, $F_{guard}$ = 21.6 kV/cm). The dashed line is a guide to the eye. (b) Lowest eigenstates and related wave functions in the CQW system calculated with nextnano[3] for F = 22.7 kV/cm. The energies of the heavy hole – conduction band and light hole – conduction band transitions agree very well with the resonance marked by the arrow in (a).

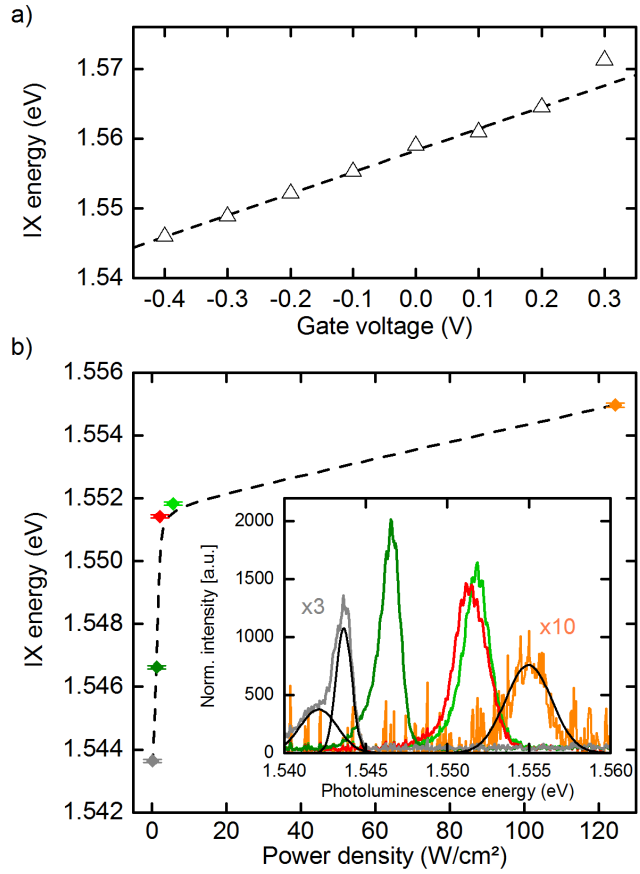

*Figure 4: (color online) (a) Quantum-confined Stark effect in the investigated field-effect device in trap configuration with $F_{guard} = 13.5$ kV/cm and $P = 8.5$ W/cm². The linear fit (dashed line) yields a coefficient of 30.9 ± 0.7 meV/V. (b) Blue-shift of the IX recombination energy as a function of laser power density. Inset shows the respective spectra using the same color code as in in the main graph. Experimental parameters are $F_{trap} = 27.0$ kV/cm, $F_{guard} = 18.9$ kV/cm, and $\hbar\omega_{laser} = 1.6095$ eV.*

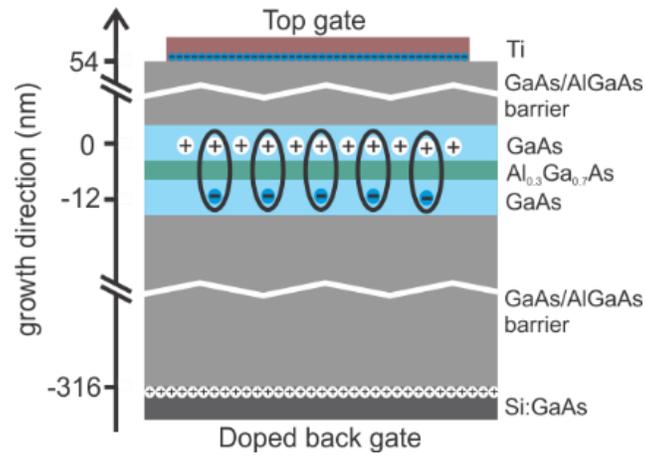

Figure 5: (color online) Sketch of the layer structure of the active CQW device along the growth direction. The coexisting 2D hole system and dipolar exciton ensemble are included in the cartoon as well as the free carriers in the titanium (Ti) top electrode and the highly n-type Si:GaAs back electrode.